\documentclass[twocolumn,showpacs,preprintnumbers,amsmath,amssymb]{revtex4}

\usepackage{oldgerm}
\usepackage{graphicx}
\usepackage{dcolumn}
\usepackage{bm}
\newcommand{\beq}{\begin{equation}}
\newcommand{\eeq}{\end{equation}}
\newcommand{\eqn}[1]{\label{eq:#1}}
\newcommand{\Eq}[1]{Eq.~(\ref{eq:#1})}

\newcommand{\bfn}{{\bf n}}
\newcommand{\bfo}{{\mathbf 1}_s}
\newcommand{\bfp}{{\bf p}}
\newcommand{\be}{\bf \hat e_0}
\newcommand{\bei}{\bf \hat e_i}
\newcommand{\scat}{a}
\newcommand{\mybar}[1]%
        {\kern 0.6pt\overline{\kern -0.6pt#1\kern -0.6pt}\kern 0.6pt}
\newcommand{\CA}{{\cal A}}
\newcommand{\CP}{{\cal P}}
\newcommand{\CV}{{\cal V}}
\newcommand{\vev}[1]{\langle #1 \rangle}

\begin{document}

\preprint{INT-PUB 03-15, MIT-CTP 3402}

\title{
 A lattice theory for low energy fermions at nonzero chemical
  potential }

\author{Jiunn-Wei Chen}%
 \email{jwc@phys.ntu.edu.tw}
\affiliation{Department of Physics and NCTS at Taipei, National Taiwan University,
Taipei 10617, Taiwan\\
and Department of Physics, CTP, Massachusetts Institute for Technology,
Cambridge, MA 02139, USA}

\author{David B. Kaplan}
 \email{dbkaplan@phys.washington.edu}
\affiliation{%
 Institute for Nuclear Theory, University of
  Washington,  
 Seattle, WA 98195-1550, USA
}%



\begin{abstract}
We construct a lattice theory describing a system of interacting
nonrelativistic  spin $s=\frac{1}{2}$ fermions at nonzero
chemical potential.  The
theory is applicable whenever the interparticle separation is large
compared to the range of the two-body potential, and does not suffer
from a sign problem.  In particular, the theory could be useful in
studying the thermodynamic limit of fermion systems for which the scattering length is much
larger than the interparticle spacing, with applications to realistic
atomic systems  and dilute neutron gases.
\end{abstract}

\pacs{71.10.Fd,74.20.Fg,03.75.Kk}
\maketitle

\section{\label{sec:1} Introduction}

In this Letter we consider dilute fermion systems with attractive
interactions, for which the effective range for two-body 
scattering is much less than the inter-particle spacing.  Such systems
are characterized by a   dimensionless number $\eta = ( a n^{1/3})$, where
$a$ is the two-body scattering length, and $n$ is the density. Such
systems are well known to 
exhibit fascinating nonperturbative behavior. For $\eta$ small and
negative (weak attraction) 
one finds the BCS solution with pairing and superconductivity. With
$\eta$ small and positive, corresponding to
strong attraction with a two-body bound state well
below  threshold, the bound pairs will Bose-Einstein condense.  In
each of these cases the behavior is nonperturbative, but since the
effective interaction is weak, the system can be successfully
described in a mean field approximation.  On the other hand, dilute
fermion systems
for which  the parameter $|\eta|$ is large are not amenable to
analytical treatment. Physical realizations include both dilute neutron
gases (the neutron-neutron scattering length is more than an order of
magnitude greater than its effective range) as well as cold, dilute
gases of fermionic atoms tuned to be near a Feshbach resonance
\cite{Inouye1998,Courteille1998,Roberts1998}.  Recently there has been intense
interest in exploring such systems experimentally
\cite{PhysicsToday2003}. In the 
limit that $|\eta|\to \infty$ one expects to see universal behavior,
so that the same dimensionless physical constants will apply equally
to the atomic and nuclear systems. 

Because of the prospects of
exploring ultracold fermionic atoms at large $|\eta|$ experimentally,
 it is of great interest to understand such systems theoretically,
 which necessitates numerical studies.
A recent numerical study of dilute fermions at large
$|\eta|$ is found in Ref.~\cite{Carlson2003}.  The authors of that
work used the
fixed-node Diffusion Monte Carlo method to extract the parameter $\xi$, defined
as the energy per particle relative to the value for a noninteracting
Fermi gas.  Since this approach is variational, the result $\xi=0.44\pm.01$  obtained
is an upper bound on the true quantity.

 The
calculation \cite{Carlson2003} is performed at fixed particle number
$N$, up to $N=38$. It is desirable to have a numerical approach which is not
variational, and which can probe thermodynamic properties of dilute
fermion gases, such as the critical temperature for the pairing
transition.
In this Letter we propose such a method, in which the fermions live on
a spacetime lattice at nonzero chemical potential. We exploit the fact
that all relevant scattering information at low density is contained in the
scattering length, so that the complicated two-body interactions can
all be replaced by an effective field theory with a single attractive
four-fermion interaction.   While lattice formulations of interacting
fermions at finite density typically encounter
a  ``sign problem'', we show that is not the case here.  There is
another issue that requires attention, however, relating to possible
instabilities introduced by a purely attractive interaction.  We
perform an analytical calculation that suggests such 
instabilities may be avoided.

\section{\label{sec:2} The lattice formulation of the effective theory}

To construct a lattice version of the continuum theory, we first
analytically continue to Euclidean spacetime, represented by
 an $N_\tau \times N_s^3$ lattice, where $N_\tau$ and $N_s$ are the number of sites
in the time and space directions respectively.  We set $\hbar=1$ and
measure all  dimensionful quantities in units of the lattice spacing,
which is taken to be the same in the space and time directions.
  We impose periodic
boundary conditions in the space directions and anti-periodic boundary
conditions in the time direction.  This allows the Euclidean path
integral 
to be interpreted as the finite temperature partition function, with
inverse temperature $\beta = N_\tau   $. The fermion fields at
site $\bfn$ are denoted by $\psi_\bfn$ and $\mybar\psi_\bfn$, and are
independent, two-component Grassmann spinors.


The continuum Euclidean action for free fermions at nonzero chemical 
potential
 is 
$\int d^4x \,\mybar
\psi\left(\partial_\tau - {\nabla^2/2M} - \mu\right)\psi$.  The corresponding lattice
action is 
$
S_0 = \sum_{\bfn} \mybar \psi_\bfn  (K_0\psi)_\bfn$, where
\beq
 (K_0\psi)_\bfn
\equiv \left(\psi_\bfn - e^{\mu  }\psi_{\bfn -
    \be}\right)
 - \sum_{i=1}^3  \frac{\left(\psi_{\bfn + \bei} - 2
    \psi_\bfn + \psi_{\bfn-\bei}\right)}{2M}
\eeq
In these expressions ${\bf \hat e_\alpha}$ is a unit vector in the $\alpha$
direction ($\alpha=0$ corresponding to Euclidean time $\tau$), $M$ is the fermion mass, and $\mu$ is the chemical
potential which is treated like the time component of an imaginary
gauge field.  This follows the
prescription in ref.~\cite{Hasenfratz:1983ba} (except that we have the
opposite sign for $\mu$) and ensures that the free energy does not
have spurious cutoff dependence. 

The free propagator computed from $S_0$ is
 \beq
G_0(\tau,\bfp) = \frac{\xi_\bfp^{\tau}}{\left(1+\Delta_\bfp/M\right)
 \left( 1+\xi_\bfp^{N_\tau}\right)}\times \begin{cases} 1 & {\rm if\ } \tau\ge 0\cr  -\xi_\bfp^{N_\tau} &
\tau<0
\end{cases}\ ,
\eeq
where 
\beq
\xi_\bfp\equiv \frac{e^{\mu  }}{1+\Delta_\bfp/M}\ ,\quad
\Delta_\bfp = 2 \sum_{i=1}^3 \sin^2\frac{  p_i}{2}\ .
\eeq 
The time $\tau$ is an integer, corresponding to propagation by $\tau$
lattice spacings in the time direction, while the momentum components are given by
$p_i = \hat p_i
\frac{2\pi}{N_s}$, with $\hat p_i$ being  integers in the range $-N_s/2 <
\hat p_i \le N_s/2$. 

Note that for $\mu<0$, $\xi_\bfp < 1$  and in the limit $N_\tau\to
\infty$ (zero temperature) $\xi_\bfp^{N_\tau}\to 0$ and there is no
propagation backward in time,
\beq
G_0(\tau,\bfp)\, \xrightarrow[\mu<0,N_\tau\to\infty] {}\,
\frac{\xi_\bfp^{\tau}\, \, \theta(\tau)}{\left(1+\Delta_\bfp/M\right)
} \ ,
\eeq
indicating the absence of antiparticles or holes in this
nonrelativistic theory at zero density.


Fermion interactions  can be
represented by an effective field theory  with a   four-fermion
interaction the strength of which is tuned to reproduce the physical
two-body scattering length.  On our lattice we generate the
interaction by means of a 
non-propagating scalar  auxiliary field $\varphi$ coupled to the fermions.  By situating
$\varphi$ along time-links of the lattice we can eliminate fermion
loops lying on surfaces of constant Euclidean time, simplifying the
analysis. Since one is
interested in pairing correlations, it is  convenient to also
introduce a constant complex  source $J$ for fermion pairs so that one
can study the pairing transition at finite volume.  These
considerations lead us to  the
action
\beq
\begin{aligned}
&S = \sum_\bfn \Bigl[
 \mybar \psi_\bfn (K \psi)_\bfn + \frac{1}{2} m^2
\varphi_\bfn^2
+ \frac{1}{2}\left(J
   \psi_{\bfn} \sigma_2 \psi_{\bfn} + {\rm h.c.}\right)\Bigl]\ ,\cr
&(K\psi)_\bfn \equiv (K_0 \psi)_\bfn - \varphi_\bfn e^\mu
\psi_{\bfn-\be}\ .
\end{aligned}
\eqn{act}
\eeq 

An important property of this theory is that after integration over $\psi$
and $\mybar \psi$, the remaining integral over $\varphi$ has positive
semidefinite measure, so that Monte Carlo integration methods may be applied. 
This result is easy to see if the source $J$ is neglected, in which
case  integrating out the fermions yields a
factor of $\det K$.  $K$ is
trivial in spin-space, and can be written as $K=\tilde K\otimes
{\mathbf 1}_s$
where $\bfo$ is the two dimensional identity matrix acting on spinor
indices, and $\tilde K$ acts only in coordinate space.  Thus $\det
K=(\det\tilde K)^2$ which is positive semidefinite, since $\tilde K$
is a real operator.

Including the constant source $J$,
the fermionic  action may be rewritten as
$
\frac{1}{2} \Psi^T \CA \Psi\ ,
$
where  we have defined
\beq \Psi = \begin{pmatrix} \psi \cr i \sigma_2 \mybar\psi^T
\end{pmatrix}\ ,\quad 
\CA = \begin{pmatrix} 
-i J & K^\dagger \\  K & - i J^* \\
  \end{pmatrix}
\begin{pmatrix} 
i\sigma_2 & 0 \\ 0 & i\sigma_2 \\
  \end{pmatrix}\ .
\eqn{adef}
\eeq
Each block in this matrix is $2N$ dimensional, where $N$ is the number
of lattice sites; $K$  acts trivially on spinor indices, the Pauli
matrix $\sigma_2$ acts trivially on coordinate indices, and $J$ is
proportional to the identity matrix in both spaces.
Note that  $K^T=K^\dagger$ since $K$ is real.
Integrating out the fermions yields a purely bosonic theory, where the path integral
over the $\varphi$ field is weighted by $exp(-\frac{m^2}{2}\sum_\bfn
\varphi_\bfn^2)$ times  the pfaffian of $\CA$:
\beq
\text{Pf}[\CA] = \begin{vmatrix}
-i J &\tilde K^\dagger \\\tilde K & - i J^* \\
  \end{vmatrix}=\sigma\,\begin{vmatrix}
 |J|^2 + \tilde K^\dagger \tilde K\ .
\eqn{pfa}
  \end{vmatrix}
\eeq
Each block in the above expression is $N$-dimensional.  Thus $\text{Pf}[\CA]$ is real and positive semidefinite for
all real values of  $J$,  $\mu$,  and $\varphi_\bfn$, up to the irrelevant
constant sign $\sigma=(-1)^{N}$, and there is no sign
problem encountered with this lattice formulation. Similar analyses exist
for two flavor QCD  \cite{Kogut:2001na} and the NJL model
\cite{Hands:2001aq}.



We have introduced in \Eq{act}  a new parameter $m^2$, which
determines the strength of the two-body interaction.  This parameter
can be related to the two-body
scattering length $\scat$ by summing the ladder diagrams with $\varphi$
exchange between the fermions at zero external momenta, zero
temperature, infinite
volume, and zero chemical potential, $\mu\to
0^-$;  the result may be equated to $\frac{4 \pi
  \scat}{M}$ (see \cite{Kaplan:1996xu}).  The result is
\beq
\frac{m^2}{M} = -\frac{1}{4\pi\scat} + L(M)\ , 
\eqn{mval}
\eeq
where $L(M)$ 
is given by the integral
\beq
\begin{aligned}
L(M) &=
 \int
  \frac{d^3\bfp}{(2\pi)^3} \, \sum_{\tau=0}^\infty \,\left(G_0(\tau,\bfp)\right)^2
\cr &=  \frac{1}{2} \int
  \frac{d^3\bfp}{(2\pi)^3}\, \frac{1}{\Delta_\bfp + \Delta_\bfp^2/2M}
\end{aligned}
\eqn{bdef}\eeq
where  the integral is over the Brillouin zone, $|p_i|\le
\pi$.

Note that since $m^2$ must be positive, scattering
lengths satisfying $\frac{1}{a} > 4\pi L(M)$ are inaccessible
in our theory without introducing an imaginary coupling for $\varphi$
and sacrificing positivity of the measure.  These values for the
scattering length correspond to two-body   bound states
with $O( 1/M)$ binding energy  in lattice units, which are not of
physical interest in any case.

\section{\label{sec:2c} Phase structure of the lattice theory}
\label{sec:2c}

An interacting  continuum limit of our lattice theory requires that we
be able to find a point in the phase diagram where both the
interparticle spacing $n^{-1/3}$ and the
scattering length $a$ diverge in lattice units.  It does not require
that the mass $M$ vanish, however. From \Eq{mval} we see it is always
possible to tune the parameter $m^2$ so that $\scat$ diverges.  What
is less obvious is that we can then choose $\mu$ so that the particle
density $n$ becomes arbitrarily small.   In general one
would expect to find a line of first order phase transitions in the
$(M,\frac{1}{a})$ plane for sufficiently large values of $\frac{1}{a}$,
corresponding to a strongly attractive four-fermion interaction.  This
follows from the well-known result  in
the continuum  that a fermion system at fixed chemical
potential with a purely
attractive interaction has no ground state, being unstable against the
formation of infinitely dense matter. On the lattice, where the maximum attainable
density is limited, one would expect that at $\mu=0$ the ground state
of the system would jump from $n=0$ to $n=O(1)$ in lattice units for
a sufficiently strong attraction.  The existence of a continuum limit
then becomes 
 whether for any mass $M$ the  $n=0$ phase at $\mu=0$ persists when $1/a\to 0$.

Therefore the first thing to compute in lattice theory is the relation
between the chemical potential $\mu$ and the fermion density $n$.  For
$n$ we take
\beq
\begin{aligned}
n &=\vev{\mybar\psi_\bfn  e^\mu \psi_{\bfn-\be}} = m^2
\vev{\varphi_\bfn} \cr
&= \frac{\int [d\varphi]\  \CP(\varphi)\  (m^2
  \varphi_\bfn)}{\int [d\varphi]\  \CP(\varphi)}
\ .
\eqn{nval}\end{aligned}
\eeq
where  $\CP(\varphi) =\left( e^{-\frac{m^2}{2}\sum_\bfn \varphi_\bfn^2}\, {\rm
  Pf}[\CA]\right)$,  and ${\rm Pf}[\CA]$  is given in \Eq{pfa}. This
definition of $n$ is more convenient than  $\partial\ln Z/\partial
\mu$, and the two definitions agree when either expression is small.

To
compute the density $n$ requires a full-scale unquenched simulation, which is
beyond the scope of this paper.  However it is instructive to
determine $n$  in the leading semi-classical (mean-field) approximation.  We
compute the effective potential obtained by integrating out the
fermions, assuming a classical background 
$\vev{\varphi}$ field;  we then determine $ \vev{\varphi}$ to be the value
that minimizes the effective potential.

Setting $J=0$ in \Eq{pfa} (and dropping the sign $\sigma$) we can 
compute ${\rm Pf[\CA]}$ exactly for constant
$\varphi$:
\beq
\begin{aligned}
{\rm Pf}[\CA] &= \det \tilde K^\dagger \tilde K \cr
&= \prod_{\{\hat p_i\}} \left[\left(1+\Delta_\bfp/M\right)^{N_\tau} + e^{N_\tau
    \mu}(1+\varphi)^{N_\tau}\right]^2
\cr
&
\equiv e^{-N_\tau N_s^3 \CV_1(\varphi)}\ ,\\
\CV_1(\varphi)&=-\frac{2}{N_s^3}\sum_{\{\hat p_i\}}\left[\ln\left(1+\frac{\Delta_\bfp}{M}\right)
      +\frac{\ln\left(1+(\xi_\bfp(1+\varphi))^{N_\tau}\right)}{N_\tau}\right].
\end{aligned}\eeq
In the $N_\tau,N_s\to\infty$ limit (large volume, zero temperature)
this becomes
\beq
\begin{aligned}
\CV_1(\varphi)= -2\int_{BZ} \frac{d^3
  \bfp}{(2\pi)^3}\, &\Biggl[\ln(1+\Delta_\bfp/M) \cr
 + &\ln\Bigl(\xi_\bfp
      (1+\varphi)\Bigr)\,\theta\Bigl(\xi_\bfp (1+\varphi)-1\Bigr)\Biggr]\ .
\eqn{effpot}\end{aligned}
\eeq
The momentum integration in the above expression is over the
Brillouin zone, $p_i\in (-\pi,\pi]$. The first term in $\CV_1$ is just the contribution from free fermions;
the second term is the quantum correction to the tree level potential,
$\CV_0(\varphi) = \frac{1}{2} m^2 \varphi^2$.  In
the mean-field approximation the
expectation value $\vev{\varphi}$ is  given by the minimum of the
effective potential
$\CV_{\rm eff}(\varphi)=(\CV_0 + \CV_1)$. 

\begin{figure}
\includegraphics[width=2.8in]{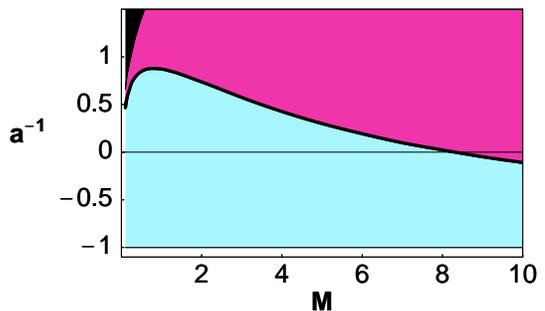}
\caption{ The density phase diagram in the
  $(M,a^{-1})$ plane at $\mu=0$ in the mean-field approximation for
  $M\ge 0.1$. The curve corresponds to a first
  order phase transition in the expectation value of the fermion
  number density $n$, with $n=0$ in the lower region and $n=O(1)$ in
  the upper region. The desired continuum physics only exists in the
  lower phase. The black region in the upper left corner is
  inaccessible for real scalar-fermion coupling.}
\label{fig:phase}
\end{figure}

The effective potential  yields the $\mu=0$  phase
diagram displayed in Fig.~\ref{fig:phase}. The horizontal axis is the
fermion mass $M$, while the vertical axis is the inverse scattering
length, $a^{-1}$.  
Moving  up the vertical axis corresponds to
  increasing the strength of the two-body attraction, and
  $a^{-1}\ge0$ describes systems with a two-body bound state. The
  curve corresponds to a jump in the fermion density from zero in the
  lower region, to $O(1)$ in the upper region.  We see that there is a
  region far from the phase transition  for $M\sim 1$ that allows small
  $a^{-1}$ of either sign. Thus one may expect to be able to perform
  simulations of fermions at finite density, where both the
  interparticle spacing $n^{-1/3}$
   and scattering length $|a|$ are large compared to the lattice
  spacing, without requiring $|\eta|$ to be small.  

\begin{figure}
\includegraphics[width=2.5in]{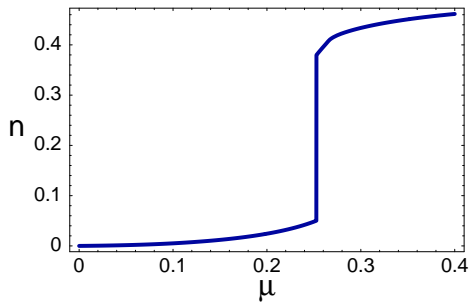}
\caption{ Density $n$ as a function of $\mu$ for $M=1$ and $a\to
  \infty$. This figure exhibits a first order phase transition as a
  function of $\mu$.  The small $\mu$ phase is where a continuum limit
is defined.}
\label{fig:density}
\end{figure}

If one were to plot the chemical potential $\mu$ along a third axis of the phase diagram,
then the critical line we have displayed would become a critical surface.  In
Fig.~\ref{fig:density} we show the fermion number density $n$ plotted
as a function of $\mu$, for fixed $M=1$ and $a^{-1}=0$.  One sees that
there is a regime where $n$ grows slowly with $\mu$, but that at a
critical chemical potential, the fermion density jumps discontinuously
as the trajectory crosses the critical surface.  We see that at least
in the mean-field approximation, there is a large region in parameter
space where we can define continuum theories with any value of $\eta\equiv (a n^{1/3})$
we desire.

One of the first
  tasks of a full-fledged simulation will be to determine the analog
  of Fig.~\ref{fig:phase} beyond the mean field approximation by
  evaluating the expression in \Eq{nval} and to show
  that a region in parameter space with an interesting continuum limit
  persists.
 An
interesting calculation to pursue subsequently  would be to
map out the critical temperature for the superfluidity phase
transition  with order parameter $\vev{\psi\psi}$ as a
function of the dimensionless quantity $\eta$. To do this one would
need to simulate the system at finite source $J$, and extrapolate to
the  $J= 0$, infinite volume limit.

\section{Discussion}
\label{sec:3}

There are numerous ways in which the lattice theory we have presented
may be modified or extended without sacrificing its crucial property
of positivity.  Several obvious possibilities are: (i) to include
a kinetic term for the $\varphi$ field, allowing for attractive fermion
interactions extended in space (for example, a
Yukawa interactions for the simulation of neutron matter, for which
$p$-wave interactions are expected to be important \cite{Bedaque:2003wj}); (ii) the $\varphi$ field could be given
derivative and spin-dependent couplings, perhaps allowing one to include the contributions from
the two-body effective range or scattering in higher partial waves; (iii) more flavors can be
introduced, which could provide an opportunity to study the importance
of three-body contact interactions and the possibility of renormalization
group limit cycles \cite{Bedaque:1998kg,Beane:2000wh,Glazek:2002hq,Bawin:2003dm}.


\begin{acknowledgments}
We wish to thank P. Bedaque, G. Bertsch, E. Braaten, J. Carlson, R. Furnstahl,
S. Reddy, S. Sharpe, D. Son  and L. Yaffe for
helpful communications. J.W.C. thanks the INT at the U. of Washington and the INFN at
the U. of Rome La Sapienza for hospitality.
J.W.C. is supported in part by the National Science Council of ROC and
the U.S. department of energy under grant DOE/ER/40762-213. D.B.K  was  supported in part by DOE grant DE-FGO3-00ER41132.
\end{acknowledgments}

\bibliography{dense}
\end{document}